\documentclass[11pt]{article}

\RequirePackage{amsthm,amsmath,amsfonts,amssymb}
\RequirePackage{graphicx}

\usepackage{natbib}
\usepackage{bm}
\usepackage{mathrsfs}
\usepackage{bbm}
\usepackage{booktabs}
\usepackage{xcolor}
\usepackage[colorlinks=true, allcolors=blue]{hyperref}
\usepackage{xspace}
\usepackage[left=3.3cm, right=3.3cm, top=3cm]{geometry}

\def\spacingset#1{\renewcommand{\baselinestretch}%
	{#1}\small\normalsize} \spacingset{1}

\setcitestyle{authoryear,open={(},close={)}}

\usepackage{authblk}
\title{\normalsize{\bf{IDENTIFYING THE RECURRENCE OF SLEEP APNEA USING A HARMONIC HIDDEN MARKOV MODEL}}}
\author[1]{Beniamino Hadj-Amar}
\author[1]{B{\"a}rbel Finkenst{\"a}dt}
\author[2]{Mark Fiecas}
\author[3]{Robert Huckstepp}
\affil[1]{\itshape\small Department of Statistics, University of Warwick}
\affil[2]{\itshape\small  Division of Biostatistics, University of Minnesota}
\affil[3]{\itshape\small  School of Life Sciences, University of Warwick \vspace{0.5cm}}
\date{March 2021}
\affil[ ] {\textcolor{blue}{\textit{Beniamino.Hadj-Amar@rice.edu}}, \textcolor{blue}{\textit{B.F.Finkenstadt@warwick.ac.uk}}}
\affil[ ] {\textcolor{blue}{\textit{mfiecas@umn.edu}}, \textcolor{blue}{\textit{R.Huckstepp@warwick.ac.uk}}}
\begin{document}

\spacingset{1.42} 

\maketitle

\begin{abstract} We propose to model time-varying  periodic and oscillatory processes by means of  
	a hidden Markov model where the states are defined through the spectral properties of a periodic regime.
	The number of states is unknown along with the relevant periodicities, the role and number of which may vary across states. 
	We address this inference problem  by a Bayesian nonparametric hidden Markov model  assuming a  sticky hierarchical Dirichlet process for 
	the switching dynamics between different states while the 
	periodicities  characterizing each state are explored by means of a  trans-dimensional Markov 
	chain Monte Carlo sampling step. We develop the full Bayesian inference algorithm and illustrate the use of our proposed methodology for  different simulation studies as well as an application related to respiratory research which focuses on the detection of apnea instances in  human breathing traces. \end{abstract}

\noindent %

\spacingset{1.45} 

\section{Introduction}

Statistical methodology for identifying periodicities  in time series can provide meaningful information about the underlying physical process. Non-stationary behavior seems to be the norm rather than the exception for physiological time series as  time-varying periodicities and other forms of rich dynamical patterns are commonly observed  in response to external perturbations  and pathological states. For example, body temperature and rest activity might exhibit changes in their periodic patterns as an individual experiences a disruption in its circadian timing system \citep{krauchi1994circadian, komarzynski2018}. Heart rate variability and electroencephalography are other examples of data that are often characterized by  time-changing spectral properties, the quantification of which can provide valuable information about the well-being of a subject \citep{malik1996heart, west1999evaluation, cohen2014analyzing, bruce2016adaptive}. This paper is motivated by modeling airflow trace data obtained from a sleep apnea study, where our objective is to identify and model the recurrence of different periodicities, which are indicative of the apneic and hyponeic events.


\subsection{A Case Study on Sleep Apnea in Humans} 
\begin{figure}[htbp]
	\centering
	\centerline{\includegraphics[height =6.9  cm, width = 15.1cm]{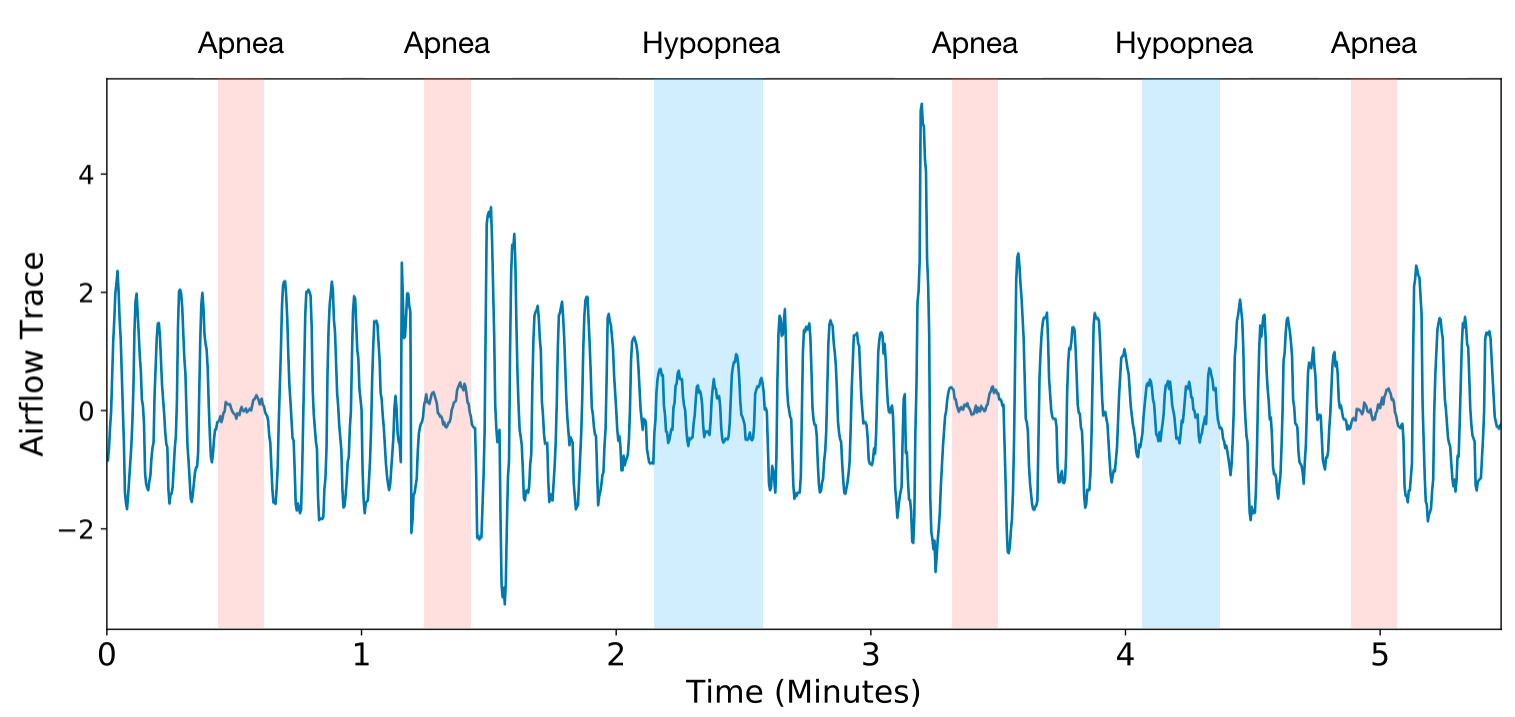}}
	\caption{Airflow trace collected over a period of five and half minutes of continuous breathing where instances of  simulated apnea and hypopnea (highlighted on the graph) were recurring over time. }
	\label{fig:human_breathing_trace_intro}
\end{figure}

Our study focuses on \textit{sleep apnea} \citep{heinzer2015prevalence}, a chronic respiratory disorder characterized by recurrent episodes of temporary ($\geq$2 breaths) cessation of breathing during sleep (about 10 seconds in humans).  Sleep apnea negatively affects several organ systems, such as the heart and kidneys in the long term. It is also associated with an increased likelihood of 
hypertension, stroke, several types of dementia, cardiovascular diseases, daytime sleepiness, depression  and a diminished quality of life 
\citep{ancoli1991dementia,teran1999association, peker2002increased, young2002epidemiology,  yaggi2005obstructive,   cooke2009sustained, dewan2015intermittent}. 
Instances of sleep apnea can be subclassified based on the degree of reduction in airflow to the lungs whereby \textit{apneas} are classified as a 
reduction of airflow by 90\% and \textit{hypopneas} require a reduction in airflow by at least 30\% (with a reduction of blood oxygen levels by at least 3\%). For example, the airflow trace shown in Figure \ref{fig:human_breathing_trace_intro}  was  collected from a human over a time span of 5.5 minutes of continuous 
breathing. During this time, apneic and hyponeic events were simulated; apneas appear in the first and second minute and around the start of the fifth minute, where there are two instances of hypopneas in the first half of the second minute and at the start of the fourth minute as marked in Figure \ref{fig:human_breathing_trace_intro}. Note these events were classified by eye by an experienced experimental researcher.   Detecting  apneic and hyponeic events during sleep  is one of the primary interests of researchers and clinicians working in the field of sleep medicine and relevant healthcare \citep{berry2017aasm}. Manual classification is a time-consuming process, and hence there is a need of a data-driven approach  for the automated classification of  these types of events. 


\subsection{Hidden Markov Models and Spectral Analysis}
Approaches to spectral analysis of nonstationary processes were first developed by \citet {priestley1965evolutionary} who introduced the concept of \textit{evolutionary spectra}, namely  spectral density functions that are time-dependent as well as localized in the frequency domain. This modeling framework was formalized as a class  
of nonstationary time series called \textit{locally stationary}  
\citep{dahlhaus1997fitting}.  Locally stationary processes can be well approximated by \textit{piecewise stationary} processes and several authors proposed to model the time-varying spectra of locally stationary time series through the piecewise constant spectra of the corresponding stationary segments \citep{ adak1998time, ombao2001automatic, davis2006structural}. 
This framework was extended to a Bayesian setting by \citet{rosen2009local} and  \citet{rosen2012adaptspec} who estimated a  time-varying spectral density using a fixed number of smoothing splines and approximated the likelihood function via a product of local Whittle likelihoods \citep{whittle1957curve}. 
Their methodology  is based on the assumption that the time series are piecewise stationary, and the underlying spectral density for each partition is smooth over frequencies.
In order to deal with changes in spectral densities with sharp peaks which can be observed for some physiological data sets such as respiratory data,  
\citet{hadj2018bayesian} proposed  a change-point analysis where they
introduced a Bayesian methodology for inferring change-points along with 
the number and values of the  periodicities affecting each segment.   While these  approaches allow us to  analyse the spectral changing properties of a process 
from a retrospective and exploratory point of view,  in order to develop a more comprehensive understanding of the process driving the data,  further modeling assumptions are needed that  quantify the  probabilistic rules governing the 
transitions as well as recurrence of different oscillatory dynamic patterns.    For example, in the context of  experimental sleep apnea research,
both, correctly classifying  the states of apnea as well as quantifying their risk of recurrence,  possibly in the context of real-time monitoring of patients,  
is of major interest to the development of treatments for breathing disorders.  

Here, we address the switching dynamics between different oscillatory states in the framework of a hidden Markov model (HMM) that assumes a
discrete latent state sequence whose transition probabilities follow a Markovian structure  (see e.g.  \citet{rabiner1989tutorial, ephraim2002hidden, cappet}). Conditioned on the state sequence, the observations are 
assumed to be   independent and  generated from a family of probability distributions, which hereafter we refer to as the \textit{emission distributions}. HMMs are arguably among the most popular statistical approaches used for modeling time series data when the observations exhibit nonstationary characteristics that can be represented by an underlying and unobserved hidden process.  These modeling approaches, also known as  hidden Markov processes and Markov-switching models, became notable by the work of \citet{baum1966statistical} and \citet{baum1967inequality}, and 
HMMs have since been successfully used  in many different applications 
\citep{krogh1994hidden, yau2011bayesian, langrock2013combining, yaghouby2015quasi,  huang2018hidden}.

As we are interested in modeling the recurrence of periodicities in the airflow trace data  we propose a harmonic HMM where the discrete latent  state sequence reflects the time-varying changes as well as recurrence of      periodic regimes as defined by their spectral properties. Furthermore, 
we pursue a  flexible nonparametric specification  within a Bayesian approach by assuming the infinite-state hierarchical Dirichlet process (HDP) as a building block \citep{teh2006}.  
The HDP-HMM approach places a Dirichlet process (DP) prior on the Markovian transition probabilities of the system, while allowing the atoms associated 
with the state-specific conditional DPs to be shared between each other, yielding an HMM with a countably infinite number of states.  The HDP-HMM therefore not only provides a  nonparametric specification of the transition distributions but also removes the need for specifying a priori the number of states. In our case study, while it is true that by looking at Figure \ref{fig:human_breathing_trace_intro} there seems to be a substantial difference between the apnea-states and normal breathing (i.e. neither apnea or hypopnea), it is also conceivable that normal breathing may exhibit many distinct periodic patterns, both with respect to  this subject  and possibly across different individuals. Yet we are interested in classifying, in an unsupervised fashion, the occurrence of apnea/hypopnea states, it might also be necessary to model states corresponding to normal breathing and other aspects that characterize respiration, such as a sigh for example.   \citet{paz2013acute} reported at least 13 forms of breathing patterns including forms of apnea,  further justifying the need to not pre-specify the number of hidden states and hence using a more versatile HDP for this application. We focus on the 
\textit{sticky} HDP-HMM by \citet{fox2011sticky},  where an additional parameter is introduced to promote self-transition with the effect that the sticky HDP-HMM more 
realistically explains the switching dynamics between states that exhibit some temporal mode  persistence. We hence extend the Bayesian methodology for the sticky HDP-HMM  
to a spectral representation of the states where
inference for the variable dimensionality regarding the number of periodicities that characterize the emission distributions of the states is achieved 
by developing an appropriate form of trans-dimensional MCMC sampling step \citep{green1995reversible}.

This article introduces a dynamic oscillatory model that is remarkably flexible while being developed in a framework that is still  computationally accessible. To the best of our knowledge, it is the first statistical methodology that exploits an HMM for analyzing the spectral properties of a time series while quantifying the probabilistic mechanism governing the transitions and recurrence of distinct dynamic patterns. The rest of the paper is organized as follows. Section \ref{sec:model} presents the model and the general framework of our Bayesian approach. 
Section \ref{sec:inference} and \ref{sec:simulation_studies} provide the inference scheme and simulation studies to show the performance 
of the proposed method. In Section \ref{sec:case_study}, we illustrate the use of our approach to detect instances of apnea in human breathing traces.


\section{A Sticky HDP-HMM with Oscillatory Emissions} \label{sec:model}

We propose a Bayesian approach relevant for analyzing observations of oscillatory dynamical systems based on an HMM. The observational model follows  \citet{andrieu1999joint} and \citet{hadj2018bayesian}, where the state-dependent data generating process is expressed via a Gaussian harmonic regression with an unknown number of periodicities. We integrate this methodology
with the nonparametric HDP-HMM model introduced by  \cite{teh2006}, where the rows of the infinite-dimensional transition matrix are framed to enable linkage between the probabilities associated with each hidden state in a hierarchical manner. Temporal mode persistence of the latent state sequence is achieved using the sticky HDP-HMM formulation of \citet{fox2011sticky}. 

Let  $\bm{y} = (y_1, \dots, y_T){'}$ be a realization of a time series whose oscillatory behavior may switch 
dynamically over time and let $\bm{z} = (z_1, \dots, z_T)'$ denote the hidden discrete-valued states of the Markov chain that 
characterize the different periodic regimes, where $z_t$ denotes the state of the Markov chain at time $t$. 
Any observation $y_t$ given the state $z_t$, is assumed to be conditionally independent of the observations and 
states at other time steps \citep{rabiner1989tutorial}. Here, a highly flexible nonparametric approach is postulated by 
assuming that the state space is unbounded, i.e. has infinitely many states as in \citet{beal2002infinite} and \citet{teh2006}. 
Thus, the Markovian structure on the state sequence $\bm{z}$ is given by 
\begin{equation} \label{eq:HMM}
z_{\,t} \, | \, z_{\, t-1}, \, ( \,\bm{\pi}_{\,j} \,)_{\,j=1}^{\infty} \,  \sim \, \bm{\pi}_{\,z_{\, t-1}}, \quad t = 1, \dots, T, 
\end{equation}
where $\bm{\pi}_{\,j} = (\pi_{j1}, \pi_{j2}, \dots)$ represents the (countably infinite) state-specific vector of transition probabilities, 
and in particular $\pi_{\,jk} = p \, (z_{\, t} = k \, | \, z_{\, t-1} = j \,)$, where  $p \, (\, \cdot \, )$ is used as a generic notation for  
probability density or  mass function, whichever appropriate. 
We assume that the initial state  has distribution $\bm{\pi}_{\,0} = (\pi_{01}, \pi_{02}, \dots)$, namely $z_{\, 0} \sim \bm{\pi}_{\,0}$.

Next, assume that each state $j$ represents a periodic regime  that is characterized by $d_j$ relevant periodicities whose frequencies 
are denoted by $\bm{\omega}_{\, j} = (\omega_{j 1}, \dots, \omega_{j \,d_j})^{'}$, recalling that periodicity is the inverse of frequency. 
Let $\bm{\beta}_{\,j } = ( \, \bm{\beta}_{j 1}^{\,'}, \dots, \,  \bm{\beta}_{j \,d_{j}}^{\,'})^{\,'}$ be the vector of linear coefficients that can 
be associated with the amplitude and phase corresponding to each frequency $\omega_{jl}$ that is of relevance to state $j$, 
where  $\bm{\beta}_{j l} = (\, \beta_{j l}^{\, (1)}, \, \beta_{j l}^{\, (2)} \, )^{'}$ and $l = 1, \dots, d_j$. 
Furthermore, let us define $ \bm{\theta}_{\, j} = (\,d_j, \, \bm{\omega}_{j}^{'} , \, \bm{\beta}_{j}^{\,'}, \sigma^2_j)^{\, '} $, where $\sigma^2_j$ 
accounts for a state-specific variance. Then, each observation is assumed to be generated from the following  emission distribution



\begin{equation} \label{eq:emission_distribution}
y_t \, \big| \,  z_{\,t} = j, \, \big( \, \bm{\theta}_{\, j} \,)_{j=1}^{\infty} \,  \sim  \, \mathcal{N} \, \Big(\,  f_{\, t j\,},  \, \sigma_{j}^{2} \, \Big), \quad t = 1, \dots, T,
\end{equation}  
where the mean function $f_{\, t j\,}$ for state $j$ at time $t$ is  \citep{andrieu1999joint, hadj2018bayesian} specified to be oscillatory, i.e.,
\begin{equation} \label{eq:oscillatory}
f_{\, t j} \, = \bm{x}_t \, \big( \bm{\omega}_{ j} \big)^{\, '} \,  \bm{\beta}_{\, j},
\end{equation}
and  the vector of basis functions $\bm{x}_t \, \big( \, \bm{\omega}_j \, \big)$ is defined as 
\begin{equation}
\label{basis_functions}
\bm{x}_t \, \big( \bm{\omega}_{j} \big) = \big(\cos(2\pi\omega_{j 1}t), \, \sin(2\pi\omega_{j 1}t), \dots, \cos(2\pi\omega_{j \, d_j}t), \, \sin(2\pi\omega_{j \, d_j}t)  \big)^{'}.
\end{equation}

The  dimension of each oscillatory function depends on the unknown number $d_j$ of  periodicities relevant to state $j$. 
Given a pre-fixed upper bound for the number of relevant periodicities per state, 
$d_{\text{max}}$, the parameter space 
$\bm{\Theta}_j$ for the vector of emission parameters $\bm{\theta_j}$  can be written as 
$ \bm{\Theta}_j = \bigcup_{d_{j} = 1}^{d_{\text{max}}} \Big\{ \, d_j \, \Big\} \times \Big\{ {\rm I\!R}^{2 d_j}  \times \bm{\Omega}_{d_j} \times {\rm I\!R}^{+}\Big\},$ 
where $ \bm{\Omega}_{d_j} = (0, 0.5)^{\, d_j}$ denotes the sample space for the frequencies of the $j$-th state. 
\citet{hadj2018bayesian} introduced this modeling approach for oscillatory data that show regime shifts in periodicity, amplitude and phase. They assume that, conditional on an unknown
number of change-points at unknown positions, the time series process can be approximated by a sequence of segments, 
each with mean functions  specified by Gaussian harmonic models of the form given in Equation \eqref{eq:oscillatory}.  Here  this approach will be combined with a nonparametric sticky HDP-HMM model \citep{fox2011sticky} which provides a structure for modeling switching dynamics and connectivity between different states.

\subsection{A Bayesian Nonparametric Framework for Unbounded Markov States} \label{sec:BNP_markov}

Dirichlet processes provide a simple description of  clustering processes where the number of clusters is not fixed a priori. 
Suitably extended to a hierarchical DP, this form of stochastic process provides a foundation for the design of state-space models in which 
the number of modes is random and inferred from the data. 
In contrast to classic methods that assume a parametric prior on the number of states, or use model selection 
techniques to determine the number of regimes in an HMM,  where here we follow \citet{beal2002infinite, teh2006} and \citet{fox2011sticky}, 
and assume the number of states to be unknown. We therefore do not need to pre-specify the number of hidden states, which provides a more flexible modeling framework. 
The DP may be used in frameworks where an element of the model is a discrete random variable of unknown cardinality \citep{hjort2010bayesian}. 
The  unbounded HMM (i.e., where the number of possible states is unknown)  can be seen as an infinite mixture model, 
where the mixing proportions are modelled as DPs \citep{beal2002infinite, rasmussen2002infinite, teh2006}.

The current state $z_{\,t}$ indexes a specific transition distribution $\bm{\pi}_{\,z_{\,t}}$ over the positive integers, 
whose probabilities are the mixing proportions for the choice of the next state $z_{t+1}$.  To allow the same set of next states to 
be reachable from each of the current states, we introduce a set of state-specific DPs, whose atoms are shared between each 
other \citep{teh2006}. As in  \citet{fox2011sticky} we implement the sticky version by increasing the expected probability of self-transitions.  
In particular, the state-specific transition 
distribution $\bm{\pi}_j$ follows the   HDP 
\begin{equation}  \label{eq:pi_DP}
\bm{\pi}_j \, \big| \, \eta, \, \kappa, \, \bm{\alpha} \sim  \text{DP} \, \Bigg( \, \eta + \kappa, \, \dfrac{\eta \, \bm{\alpha} + \kappa \, \delta_j}{\eta + \kappa}  \, \Bigg),
\end{equation}
where
\begin{equation*}\label{eq:GEM}
\bm{\alpha} \, \big| \, \gamma \sim \text{GEM} \, ( \, \gamma \, ).   
\end{equation*}


Here, the sequence $\bm{\alpha} = (\, \alpha_{\, k} \,)_{k=1}^{\infty}$ can be seen as a \textit{global} probability distribution over 
the positive integers that ties together the  transition distributions $\bm{\pi}_j$ and guarantees that they have the same support. 
We denote by GEM ($  \gamma  )$ \footnote{GEM is an abbreviation for Griffiths, Engen and McCloskey, see  \citet{ignatov1982constant, perman1992size, pitman1996blackwell}   for background.} the \textit{stick-breaking construction}  \citep{sethuraman1994constructive, pitman2002poisson} of $\bm{\alpha}$ as 
\begin{equation} \label{eq:stick_breaking_1}
\alpha_k = \nu_k \,  \prod_{l=1}^{k-1}  \,  ( 1 - \nu_l), 
\end{equation}
where
\begin{equation} \label{eq:stick_breaking_2}
\nu_k \, | \, \gamma \sim \text{Beta} \, (\, 1, \,  \gamma \, ),  
\end{equation}
for $  k = 1, 2, \dots  $,   and  $\gamma$ is a positive real number that controls the expected value of the number of elements in $\bm{\alpha}$ 
with significant probability mass. Equations \eqref{eq:stick_breaking_1} and \eqref{eq:stick_breaking_2} can be motivated by the 
equivalent process where a stick of  length one is split into lengths specified by the weights $\alpha_k$, where the $k^{\,  th}$ proportion 
is a random fraction $\nu_k$ of the remaining stick after the preceding $(\, k -1 \, )$ proportions have been constructed. 
The stick-breaking construction ensures that the  sequence $\bm{\alpha}$ satisfies $\sum_{k=1}^{\infty} \alpha_{\, k} = 1$ with probability one.  

Conditional on $\bm{\alpha}$, the hierarchical structure given in Equation \eqref{eq:pi_DP} indicates that the state-specific 
transition distributions $\bm{\pi}_{\, j}$ are distributed according to a DP with \textit{concentration parameter} $\eta + \kappa$ 
and \textit{base distribution} $(\eta \, \bm{\alpha} + \kappa \, \delta_j)/(\eta + \kappa)$, that is itself a DP.
Here, $\eta$ is a positive real number that controls the variability of the $\bm{\pi}_{\, j}$'s around $\bm{\alpha}$, 
while $\kappa$ is a positive real number that inflates the expected probability of a self-transition \citep{fox2011sticky}, 
and $ \delta_j $ denotes a unit-mass measure concentrated at $j$. By setting $\kappa = 0$ in Equation \eqref{eq:pi_DP}, 
we obtain the non-sticky HDP-HMM framework proposed by \citet{teh2006}. It was noted that this specification could result in  an 
unrealistically rapid alternation between different (and often redundant) states. The
\textit{sticky} formulation of \citet{fox2011sticky}  allows  for more temporal state persistence by inflating the 
expected probabilities of self-transitions  by an amount proportional to $\kappa$, i.e.

\begin{equation*}
\mathbb{E} \, \big[ \, \pi_{jk} \, | \, \eta, \, \kappa, \, \bm{\alpha} \big] = \dfrac{\eta}{\eta + \kappa} \, \alpha_k + \dfrac{\kappa}{\eta + \kappa} \, \delta \,(j, \, k),
\end{equation*}
where $\delta \,(j, \, k) = 1$  if $k = j$ and zero otherwise. Though the sticky parameter parameter increases temporal persistence of the hidden state sequence,  several simulations studies (see e.g.  \citealt{fox2011sticky}) suggest that the sticky HDP-HMM is still capable to capture and identify states that are short in duration  by inferring a small probability of self-transition.

\section{Inference} \label{sec:inference}
Our inference scheme  is formulated within a full Bayesian framework, where our proposed sampler alternates between updating the emission and the HMM parameters.  Section \ref{sec:emission_parameters} presents a reversible jump MCMC based algorithm to obtain posterior samples of the emission parameters $\bm{\theta}_j$, where a trans-dimensional MCMC sampler is developed to explore subspaces of variable dimensionality regarding the number of periodicities that characterize state $j$.  This part of the inference scheme follows a similar structure to the one  presented by \citet{andrieu1999joint} and \citet{hadj2018bayesian}. In Section \ref{sec:HMM_parameters} we address  model search on the number of states by exploiting the \textit{Chinese restaurant franchise with loyal customers} \citep{fox2011sticky}, a metaphor that provides the building blocks to perform Bayesian nonparametric inference for updating the HMM parameters. The details related to this part of the sampler follow the scheme presented in \citet{fox2011sticky}. The resulting Gibbs sampler for the full estimation algorithm is described in Section \ref{sec:gibbs_sampler} while in Section \ref{sec:label_switching} we address  the label switching problem related to our proposed approach. 


\subsection{Emission Parameters} \label{sec:emission_parameters}
Conditional on the state sequence $\bm{z}$, the observations $\bm{y}$ are implicitly partitioned into a finite number of states, 
where each state refers to at least one segment of the time series. When a type of periodic behavior recurs over time, the corresponding 
state is necessarily related to more than one segment. Let $\bm{y}^{\,*}_{j} =  ( \, \bm{y}^{\,'}_{j1}, \bm{y}^{\,'}_{j2}, \dots, \bm{y}^{\,'}_{j R_{j}})^{\, '} $ 
be the vector of (non-adjacent) segments  that are assigned to state $j$, where $\bm{y}_{jr}$ denotes the $r^{\, th}$ segment of the time series for which $z_{\,t} = j$ and $R_j$ 
is the total number of segments assigned to that state. Then, the likelihood of the emission parameter $ \bm{\theta}_j $ given  the observations in $\bm{y}^{\,*}_j$ is
\begin{equation}
\label{likelik_segment}
\mathscr{L} ( \, \bm{\theta}_{\, j} \, | \, \bm{y}^{\,*}_{j} \,  ) =  ( \, 2\pi \sigma^2_j \, )^{\,-T^{\,*}_j/2} \exp \Bigg[ -\dfrac{1}{2\sigma^2_j} \,  \sum_{ t \, \in \, I^{\,*}_j} \Bigg\{ \,  y_t - \bm{x}_t \, \big( \bm{\omega}_{ j} \big)^{\, '} \,  \bm{\beta}_{\, j} \,  \Bigg\}^{\, 2} \,\Bigg],
\end{equation} where $I^{\,*}_j$  and $T_j^{\,*}$ denote the set of time points and  number of observations, respectively, associated with   $\bm{y}_j^{\, *}$.


Following \citet{hadj2018bayesian}, we assume independent Poisson prior distributions for the number of frequencies $d_j$ for each state $j$, constrained on $1 \leq d_j \leq d_{max}$.  
Conditional on $d_j$, we choose a uniform prior for the frequencies $ \omega_{j, \, l} \sim \text{Uniform}(0, \phi_{\omega}), \, \,  l = 1, \dots, d_j$,  where $0 < \phi_{\omega} < 0.5$.  The value of $\phi_\omega$ can be chosen to be informative in the sense that it may reflect prior information about the significant frequencies that drive the overall variation in the data, for example  $\phi_\omega$ may be assumed to be in the low frequencies range $ 0 < \phi_\omega < 0.1 $.  Analogous to a Bayesian regression  \citep{Bishop:2006:PRM:1162264}, a zero-mean isotropic Gaussian prior is assumed for the  coefficients of the $j^{th}$ regime,
$\bm{\beta}_{\, j} \sim \mathcal{\bm{N}}_{2d_j} (\, \bm{0}, \, \sigma^2_{\beta} \, \bm{I}\, )$, where the prior variance $\sigma^2_\beta$  is fixed at a relatively large value (e.g., in our case $10^{\,2})$. The prior on the residual variance $\sigma^2_j$ of state $j$ is specified as
$\text{Inverse-Gamma} \, \big(\frac{\xi_0}{2}, \frac{\tau_0}{2}\big)$, where $\xi_0$ and $\tau_0$ are fixed at small values, noticing that when $\xi_0 = \tau_0 = 0$
we obtain Jeffreys' uninformative prior \citep{bernardo2009bayesian}.

Bayesian inference on $\bm{\theta}_j$ is built upon the following factorization of the joint posterior distribution
\begin{equation} \label{eq:posterior_theta_j}
p \,  ( \, \bm{\theta}_j \, | \, \bm{y}^{\,*}_{j} ) = p \, ( \, d_j \, | \, \bm{y}^{\,*}_{j} ) \,  p \, ( \, \bm{\omega}_j \, |\, d_j, \, \bm{y}^{\,*}_{j} ) \, p \, ( \, \bm{\beta}_j \, |\, \bm{\omega}_j, \, d_j, \, \bm{y}^{\,*}_{j} ) \, p \, ( \, \sigma^2_j \, |\, \bm{\beta}_j, \, \bm{\omega}_j, \, d_j, \, \bm{y}^{\,*}_{j} ).
\end{equation} Sampling from \eqref{eq:posterior_theta_j} gives rise to a model selection problem regarding the 
number of periodicities, thus requiring an inference algorithm that is able to  explore subspaces of variable dimensionality. 
This will be addressed by the reversible-jump   sampling step introduced in the following section. 

\subsubsection{Reversible-Jump Sampler}
Here we provide the details for drawing  $\bm{\theta}_j$ from the posterior distribution $ p \,  ( \, \bm{\theta}_j \, | \, \bm{y}^{\,*}_{j} )$ given in Equation \eqref{eq:posterior_theta_j}. 
Our methodology follows \citet{andrieu1999joint} and \citet{hadj2018bayesian} and is based on the principles of reversible-jump MCMC introduced in \citet{green1995reversible}.  
Notice that, conditional on the state sequence $\bm{z}$,  the emission parameters $\bm{\theta}_j$ can be updated independently and in parallel for each of the current states. 
Hence, for the rest of this subsection and for ease of notation, we drop the subscript corresponding to the  j$^{th}$ state.

At each iteration of the algorithm, a random choice with probabilities given in \eqref{eq:transition_prob} based on the current number of frequencies $d$ will dictate whether to 
add a frequency (\textit{birth step}) with probability $b_d$, remove a frequency (\textit{death step})  with probability $r_d$, or update the frequencies (\textit{within step}) with probability 
$\mu_d = 1 - b_d - r_d$, where

\begin{equation}
\label{eq:transition_prob} 
b_d = c \, \text{min} \Bigg\{ 1, \dfrac{p\, (\,d+1\,)}{p\,(\,d\,)} \Bigg\}, \qquad r_{d+1} = c \, \text{min} \Bigg\{1, \dfrac{p\,(\,d\,)}{p\,(\,d+1\,)} \Bigg\},
\end{equation} for some constant $c$ $\in [0, \frac{1}{2}]$ and   $p \, (\,d\,)$ is the prior probability.  Here, as in \citet{hadj2018bayesian}, we fixed $c=0.4$ but other values are admissible 
as long as $c$ is not larger than $0.5$ to guarantee that the sum of the probabilities  does not exceed 1 for some values of $c$.  
Naturally, $b_{d_{\text{max}}} = r_{1} = 0$. An outline of these moves is as follows (further details are provided in  Supplementary Material \textcolor{blue}{A}).

{\it Within-Model Move:} \label{segment_model_within} 
Conditional on the number of frequencies $d$, the vector of frequencies $\bm{\omega}$ is sampled, where  we update the frequencies one-at-a-time  using Metropolis-Hastings (M-H)  steps, with target distribution
\begin{equation} \label{posterior_omega}
p \, (\bm{\omega} \, | \, \bm{\beta},  \, \sigma^2, \, d, \, \bm{y}^{\,*} ) \propto \exp \Bigg[ -\dfrac{1}{2\sigma^2} \sum_{ t \, \in \, I^{\,*}} \Big\{ y_t - \bm{x}_t \, \big( \, \bm{\omega} \, \big)^{\, '} \,  \bm{\beta} \, \Big\}^{2}  \Bigg] \mathbbm{1}_{\big[ \, \bm{\omega} \,  \in  \, \bm{\Omega}_{d} \big] \, }.
\end{equation}
Specifically, the proposal distribution is a combination of a Normal random walk centred around the current frequency  
and a draw from the periodogram of $\hat{\bm{y}}$, where $\hat{\bm{y}}$ denotes a segment of data randomly chosen from  $\bm{y}^{\, *}$ 
with probability proportional to the number of observations belonging to that segment.  Naturally, when a state does not recur over time, i.e. 
when a state refers to only one segment of the time series, that segment is chosen with probability one. Next, updating the vector of linear coefficients $\bm{\beta}$ and the residual variance $\sigma^{\,2}$ is carried out as in the fashion of the usual normal Bayesian regression setting \citep{gelman2014bayesian}.
Hence,  $\bm{\beta}$ is updated in a Gibbs step  from 
\begin{equation} \label{posterior_beta}
\bm{\beta} \, \big| \,  \bm{\omega}, \, \sigma^2, \,  d, \, \bm{y}^{\,*} \sim \bm{\mathcal{N}}_{2d} \, (\, \hat{\bm{\beta}}, \, \bm{V}_{\beta}), 
\end{equation} where \begin{equation}
\begin{split}
\bm{V}_{\beta} &= \bigg( \sigma^{-2}_\beta \,  \bm{I} + \sigma^{-2} \bm{X}^{\,*}(\bm{\omega})^{\,'} \bm{X}^{\,*}(\bm{\omega}) \bigg)^{-1}, \\
\hat{\bm{\beta}} &= \bm{V}_{\beta} \,  \big( \sigma^{-2} \bm{X}^{*}(\bm{\omega})^{\,'} \bm{y}^{*}  \big),
\end{split} 
\end{equation} and we denote with $\bm{X}^{\,*}(\bm{\omega})$ the design matrix with rows given by $\bm{x}_t \, \big( \bm{\omega}$ \big) (Equation \ref{basis_functions}), for $t \in I^{\,*}$. 
Finally,  $\sigma^{\,2}$ is  drawn in a Gibbs step  directly from 
\begin{equation}
\label{inverse_gamma}
\sigma^2 \, \big| \, \bm{\beta}, \, \bm{\omega}, \, d, \, \bm{y}^{\,*} \sim \text{Inverse-Gamma} \, \Bigg( \, \dfrac{T + \xi_0}{2}, \, \dfrac{\tau_0 + \sum_{ t \, \in \, I^{\,*}}\Big\{ \, y_t - \bm{x}_t \, \big( \, \bm{\omega} \, \big)^{\, '} \,  \bm{\beta} \, \Big\}^{2}}{2}  \Bigg).  
\end{equation}

{\it Trans-Dimensional Moves: }  \label{segment_model_between} 
For these types of move, the number of periodicities is either proposed to increase by one (birth) or decrease by one (death) \citep{green1995reversible}.  
If a birth move is attempted, we have that $d^{\,p} = d^{\,c} + 1$, where we denote with superscripts \textit{c} and \textit{p}, the current and proposed values, respectively. 
The proposed vector of frequencies is obtained by drawing an additional frequency to be included  in the current vector. On the other hand if a death move is chosen, 
we have that $d^{\,p} = d^{\,c} - 1$ and one of the current periodicities is randomly selected to be deleted.  Conditional on the proposed vector of frequencies, 
the  vector of linear coefficients and the residual variance are  sampled as in the within-model move described above.  For both birth and death moves, the updates are jointly 
accepted or rejected in a M-H step.

\subsection{HMM Parameters} \label{sec:HMM_parameters}

We explain how to perform posterior inference about the probability distribution $\bm{\alpha}$, the transition probabilities $\bm{\pi}_j$ 
and  the state sequence $\bm{z}$.  The  \textit{Chinese restaurant franchise with loyal customers} presented by \citet{fox2011sticky}, 
which extends the \textit{Chinese restaurant franchise} introduced by \citet{teh2006}, is a metaphor that can be used to express the generative 
process behind the sticky version of the HDP and provides a general framework for performing inference. 
A  high level summary of the metaphor is as follows: in a \textit{Chinese restaurant franchise} the analogy of a \textit{Chinese restaurant process} 
\citep{aldous1985exchangeability} is extended to a set of restaurants, where an infinite global menu of dishes is shared across these restaurants.  
The process of seating customers at tables happens in a similar way as for the Chinese restaurant process, but is restaurant-specific. 
The process of choosing dishes at a specific table happens franchise-wide, namely the dishes are selected with probability proportional 
to the number of tables (in the entire franchise) that have previously served that dish. However, in the Chinese restaurant franchise with loyal 
customers, each restaurant in the franchise has a speciality dish which may keep many generations of customers eating in the same restaurant.

Let $y_{j1}, \dots, y_{j N_j} $ denote   the set of customers in restaurant $j$, where $N_j$ is the number of customers in restaurant $j$ and each customer is pre-allocated to a specific restaurant designated by that customer's group $j$. Let us also define indicator random variables $t_{ji}$ and $k_{jt}$, such that $t_{ji}$ indicates the table assignment for customer $i$ in  restaurant $j$, and $k_{jt}$  the dish assignment for table $t$ in  restaurant $j$. In the Chinese restaurant franchise with loyal customers, customer $i$ in restaurant $j$ chooses a table via $t_{ji} \sim  \bm{\tilde{\pi}}_j,$ where $\bm{\tilde{\pi}}_j \sim \text{GEM} \, (\eta + \kappa)$, and $\eta$ and $\kappa$ are as in Section \ref{sec:BNP_markov}.  Each table is assigned a dish via  $k_{jt} \sim (\eta \, \bm{\alpha} + \kappa \, \delta_j)/(\eta + \kappa) $, so that there is more weight on the house speciality dish, namely the dish that has the same index as the restaurant. Here,  $\bm{\alpha}$ follows a DP with concentration parameters $\gamma$ and can be seen as a collection of ratings for the dishes served in the global menu. Note that in the HMM formulated in Equation \eqref{eq:HMM}, the value of the hidden state $z_{\,t}$ corresponds to the dish index, i.e. $ k_{j t_{\!ji}} = z_{ji} =  z_{\,t}$, where we suppose there exist a bijection $f : t \rightarrow ji$ of time indexes $t$ to restaurant-customer indexes $ji$. Furthermore, as suggested in \citet{fox2011sticky}, we augment the space and introduce  \textit{considered} dishes $\bar{k}_{jt}$ and \textit{override} variables $o_{jt}$ so that we have the following generative process
\begin{equation*} \label{eq:served_considered}
\begin{split}
\bar{k}_{jt} \, | \, \bm{\alpha} &\sim \bm{\alpha} \\
o_{jt} \, | \, \eta, \, \kappa \, &\sim \text{Bernoulli}\,\bigg( \, \dfrac{\kappa}{\eta + \kappa} \, \bigg) \\
k_{jt} \, | \, \bar{k}_{jt}, \, o_{jt} \, &= \begin{cases}  \,\bar{k}_{jt}, & o_{jt} = 0, \\
\,j, & o_{jt} = 1.
\end{cases}
\end{split}
\end{equation*}
Thus, a table first considers a dish $\bar{k}_{jt}$ without taking into account the dish of the house, i.e. $\bar{k}_{jt}$ is chosen from the infinite buffet line according to the ratings provided by $\bm{\alpha}$. Then, the dish $k_{jt}$ that is actually being served  can be  the house-speciality dish $j$,  with probability $\rho = \kappa / (\eta + \kappa)$, or the initially considered dish $\bar{k}_{jt}$, with probability $ 1- \rho$.   As shown in \citet{fox2011sticky},  table counts $\bar{m}_{jk}$ of considered dishes  are  sufficient statistics for updating the collection of dish ratings $\bm{\alpha}$, where $\bar{m}_{jk}$ denotes how many of the tables in restaurant $j$ considered dish $k$. The sampling of $\bar{m}_{jk}$  is additionally simplified by introducing the table counts $m_{jk}$ of served dishes and override variables $o_{jt}$. In the next section we describe  a Gibbs sampler which alternates between updating the hidden states $\bm{z}$, dish ratings $\bm{\alpha}$, transition probabilities $\bm{\pi}_j$,  newly introduced random variables  $m_{jk}$, $o_{jt}$ and $\bar{m}_{jk}$, emission parameters $\bm{\theta_j}$, as well as the hyperparameters  $\gamma$, $\eta$ and $\kappa$.

\subsubsection{Gibbs Sampler} \label{sec:gibbs_sampler}
We follow \citet{kivinen2007learning} and \citet{fox2011sticky} and consider a Gibbs sampler  which uses finite approximations to the DP to allow  sampling in blocks of the state sequence $\bm{z}$. In particular, conditioned on  observations $\bm{y}$,  transition probabilities $\bm{\pi}_j$ and emission parameters $\bm{\theta}_j$, the hidden states $\bm{z}$ are sampled using a variant of the well-known HMM forward-backward procedure (see Supplementary Material \textcolor{blue}{B}) presented in \citet{rabiner1989tutorial}. In order to use this scheme, we must truncate the countably infinite transition distributions $\bm{\pi}_j$ (and global menu $\bm{\alpha}$), and this is achieved  using the $\,K_{\text{max}}$-limit approximation to a DP \citep{ishwaran2002exact}, i.e. $\text{GEM}_{\,K_{\text{max}}} \, ( \gamma )  := \text{Dir} \,  \big( \gamma/K_{\text{max}}, \dots, \gamma/K_{\text{max}} \big),$ where  the truncation level $K_{\text{max}}$ is a number that exceeds the total number of expected HMM states, and $\text{Dir} \, ( \cdot ) $ denote the Dirichlet distribution.  Following \citet{fox2011sticky}, conditioned on the state sequence $\bm{z}$ and collection of dish ratings $\bm{\alpha}$,   we sample the auxiliary variables $m_{jk}$, $o_{jt}$ and $\bar{m}_{jk}$ as described in Supplementary Material \textcolor{blue}{C.2}. Dish ratings $\bm{\alpha}$ and transition distributions $\bm{\pi}_j$ are then updated from the following  posterior distributions
\begin{equation*}
\label{eq:posterior_alpha_1}
\begin{split}
\bm{\alpha} \, | \, \bm{\bar{m}}, \gamma &\sim  \, \text{Dir} \, \big(\gamma/K_{\text{max}} + \bar{m}_{\cdot 1}, \dots, \gamma/K_{\text{max}}  + \bar{m}_{\cdot K_{\text{max}}}\big)\\
\bm{\pi}_j \, | \, \bm{z}, \, \bm{\alpha},\, \eta, \, \kappa \, &\sim \, \text{Dir} \, \big( \eta \, \alpha_1 + n_{j1}, \dots,  \eta \, \alpha_j + \kappa +  n_{jj}, \dots, \eta \,  \alpha_{K_{\text{max}}} + n_{j K_{\text{max}}} \big),
\end{split}
\end{equation*}
for each state $j = 1, \dots, K_{max}$. Here, $\bm{\bar{m}}$ is the vector of table counts of considered dishes for the whole franchise, and marginal counts are described with dots, so that $ \bar{m}_{\cdot k} = \sum_{j=1}^{K_{\text{max}}} \bar{m}_{jk} $ is the number of tables in the whole franchise considering dish $k$. We denote with $n_{jk}$  the number of Markov chain transitions from state $j$ to state $k$ in the  hidden sequence $\bm{z}$.  Next, given the state sequence $\bm{z}$  and transition probabilities $\bm{\pi}_j$, we draw  the emission parameters $\bm{\theta}_j$ for each of the currently instantiated  state as described in Section \ref{sec:emission_parameters}, where each reversible-jump MCMC update is run for several iterations.  We also need to update the emission parameters for states which are not instantiated (namely, those states among $\{1, \dots, K_{\text{max}} \}$ that are not represented during a particular iteration of the sampler), and hence we draw the corresponding emission parameters from their priors. For computational or modeling reasons, the latter may be also performed for those instantiated states that do not contain a minimum number of observations. Finally, we  sample the hyperparameters $\gamma$, $\eta$ and $\kappa$ in a Gibbs step (see Supplementary Material \textcolor{blue}{C.3}). 

For the HDP-HMM, different procedures have been applied for sampling the hidden state sequence $\bm{z}$. \citet{teh2006} originally introduced an approach based on a Gibbs sampler which has been shown to suffer from slow mixing behavior  due to   strong correlations that is frequently observed in the data at nearby time points.   \citet{van2008beam} presented a \textit{beam sampling} algorithm   that combines a slice sampler \citep{neal2003slice} with dynamic programming. This allows to constrain the number of reachable states at each MCMC iteration to a finite number, where the entire hidden sequence $\bm{z}$  is drawn in block 
using a form of forward-backward filtering scheme. However, \citet{fox2011sticky} showed that applications of the beam sampler to the HDP-HMM resulted in slower mixing rates compared to the forward-backward procedure that we use in our truncated model. Recently, \citet{tripuraneni2015particle} developed a particle Gibbs MCMC algorithm  \citep{andrieu2010particle} which uses an efficient proposal and makes use of ancestor sampling  to enhance the mixing rate.

\subsubsection{Label Switching } \label{sec:label_switching} The proposed approach may suffer from  \textit{label switching} (see e.g. \citet{redner1984mixture, stephens2000dealing, jasra2005markov}) since  the likelihood is invariant   under permutations of labelling of the mixture components, for both  hidden state labels $\{1, \dots, K_{max}\}$ and frequency labels $\{1, \dots, d_{max}\}$ in each state. The label switching problem occurs when using Bayesian mixture models and needs to be addressed in order to draw meaningful inference about the posterior model parameters. In our multiple model search, the frequencies  (and their corresponding linear coefficients) are identified by  keeping them in ascending order for every iteration of the sampler.   Posterior samples of the model parameters corresponding to different hidden states are post-processed (after the full estimation run) using the relabelling algorithm developed by \citet{stephens2000dealing}. The basic idea behind this algorithm is to find permutations of the MCMC samples in such a way that the Kullback-Leibler (KL) divergence \citep{kullback1951information} between the  `true' distribution on clusterings, say $P \, ( \bm{\theta} )$, and a matrix of classification probabilities, say $\bm{Q}$, is minimized. The KL distance is given by $ d( \bm{Q}, \, P \, ( \bm{\theta} )  )_{\, KL} =  \sum_{t} \sum_{j} p_{tj} (\bm{\theta}) \log \frac{p_{tj} (\bm{\theta)}}{q_{tj}}$, where $p_{tj} (\bm{\theta}) = p \, (z_t =  j \, | \, z_{t-1}, \, \bm{y}, \bm{\pi}, \bm{\theta}  \, )$ is part of the MCMC output obtained as in Supplementary Material \textcolor{blue}{C.1}, and $q_{tj}$ is the probability that observation $t$ is assigned to class $j$.   The algorithm iterates between estimating $\bm{Q}$ and the most likely permutation of the hidden labels for each MCMC iteration. We chose the strategy of \citet{stephens2000dealing} since it has been shown to perform very efficiently in terms of finding the correct relabelling (see e.g. \citet{rodriguez2014label}). However,  it may be computationally quite intensive in memory since it requires the storage of a matrix of probabilities  of dimension $N \times T \times K_{max}$, where $N$ is the number of MCMC samples. Furthermore, at each iterative step, the algorithm requires to go over $K_{max}!$ permutations of the labels for each MCMC iteration, which might significantly slow down the  computation when using large values of $K_{max}$.  Related approaches to the label switching issue include pivotal reordering algorithms \citep{marin2005bayesian}, label invariant loss functions \citep{celeux2000computational, hurn2003estimating} and equivalent classes representative methods \citep{papastamoulis2010artificial}, where an overview of these strategies can be found in \citet{rodriguez2014label}.

\section{Simulation Studies} \label{sec:simulation_studies}

This section presents results of simulation studies to explore the performance of our proposed methodology in two different settings. In the first scenario the data are generated from the model described in Section \ref{sec:model} and thus this simulation study provides a ``sanity'' check that the algorithm is indeed retrieving the correct pre-fixed parameters. We also investigate signal extraction  for the case that the innovations  come  from a heavy-tailed t-distribution instead of a Gaussian. Our second  study deals with artificial data from an HMM whose emission distributions are characterized by oscillatory dynamics generated by state-specific autoregressive (AR) time series models. Julia code that implements our  procedure is available at \href{https://github.com/Beniamino92/HHMM}{\texttt{https://github.com/Beniamino92/HHMM}}.

\subsection{Illustrative Example} \label{sec:illustrative_example_simul}
\begin{figure}[htbp]
	\centering
	\centerline{\includegraphics[scale = 0.40]{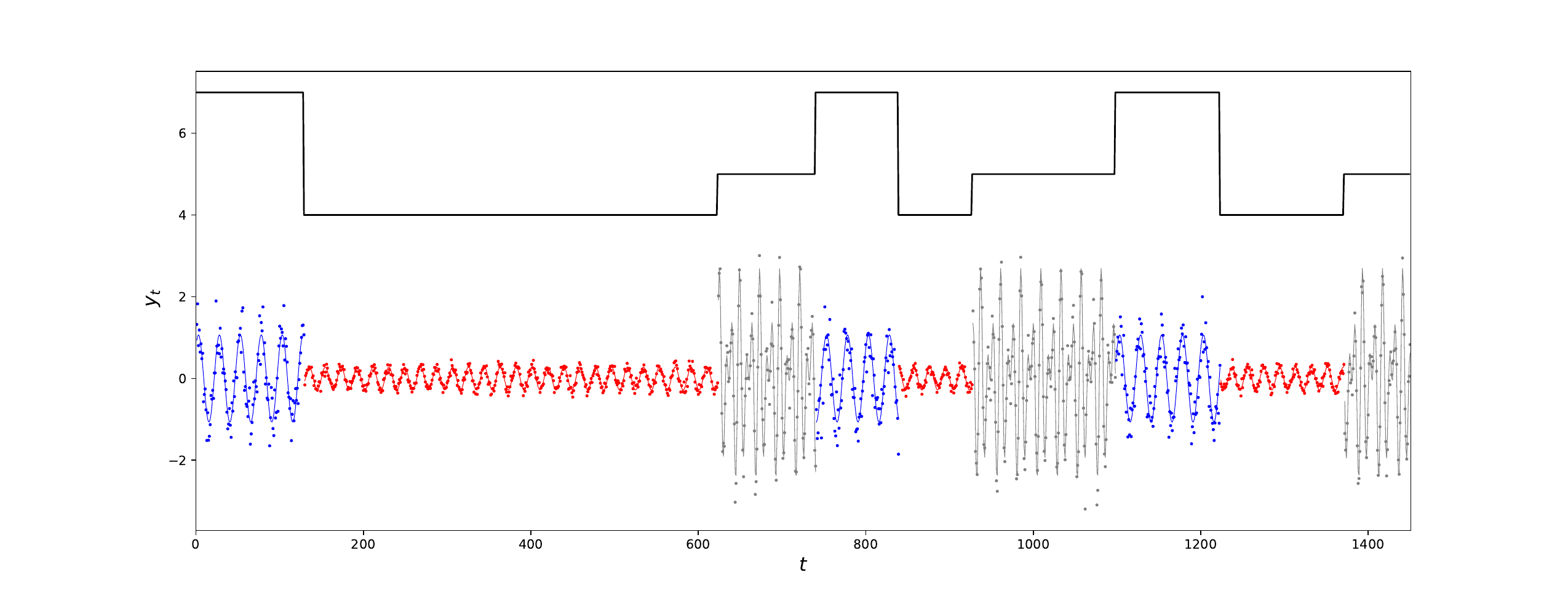}}

	\caption{Illustrative Example. Dots represent the simulated time series,  where the different colors corresponds to (true) different regimes. The state-specific estimated oscillatory mean function is  displayed as a solid curve, and the estimated state sequence as a piecewise horizontal line at the top part of the graph. }
	
	\label{fig:illustrative_example}
	
\end{figure}

We generated a time series consisting of $T = 1450$ data points  from a  three-state HMM with 
the following transition probability matrix showing high probabilities of self transition along the diagonal
and Gaussian oscillatory emissions as  specified in Equation \eqref{eq:emission_distribution}, where the parameters  of each of the three regimes and the transition probability matrix are given in Supplementary Material \textcolor{blue}{A}. A realization from this model is displayed in Figure \ref{fig:illustrative_example}. The prior mean on the number of frequencies $d_j$  is set equal to 1 and we  place a Gamma $(1, 0.01)$ prior on the concentration parameters $\gamma$ and $(\eta + \kappa)$, and a Beta $(100, 1)$ prior on the self-transition proportion $\rho$ as in \citet{fox2011sticky}. The maximum number of periodicities per regime $d_{\text{max}}$ is set to 5, while the truncation level $K_{\text{max}}$ for the DP approximation is set equal to 7. Also, we set $\phi_{\omega} = 0.25$ as a threshold for the uniform prior. The proposed estimation algorithm is run for 15,000 iterations, 3,000 of which were discarded as burn-in.  At each iteration, for each instantiated set of emission parameters, 2 reversible-jump MCMC updates were performed. The full estimation algorithm took  31 minutes with a program written in Julia 0.62 on an Intel\textsuperscript{\textregistered} Core\textsuperscript{TM} i7  2.2 GHz Processor 8 GB RAM. For our experiments, we used the R package \textit{label.switching} of \citet{papast2016} to post-process the MCMC output with the relabelling algorithm of \citet{stephens2000dealing}.

\begin{table}[htbp]
	\centering
	\caption{Illustrative example. (left panel) posterior probabilities for number of distinct states $k$; (right panel) posterior probabilities for number of frequencies in each state, conditioned on $k = 3$. }
	\label{table:posterior_k_m_illustrative_1}
	\begin{tabular}{cc}
		\hline \\[-0.9em]
		\hline \\[-0.9em]
		$k $ & $\hat{\pi} \, (\, k \, | \, \bm{y})$ \\[.1em] \hline
		1 &      0.00\\
		2 &      0.00\\
		3 &     0.99 \\
		4 &     0.01 \\
		5 &     0.00 \\ 
		6 &     0.00 \\ 
		7 &     0.00 \\ 
		\hline
	\end{tabular} \quad \quad
	\begin{tabular}{cccc}
		\hline \\[-0.9em]
		\hline \\[-0.9em]
		$m$ & $\hat{\pi} \, (\, d_1 \, | \, k = 3,  \, \bm{y})$ & $\hat{\pi} \, (\, d_2 \, | \, k = 3,  \, \bm{y})$ & $\hat{\pi} \, (\, d_3 \, | \, k = 3,  \, \bm{y})$ \\[.1em] \hline
		1 &  0.99    & 1.00     &  0.01    \\
		2 & 0.01     & 0.00     &  0.99    \\
		3 & 0.00      & 0.00     &  0.00    \\
		4 & 0.00     & 0.00     &   0.00   \\
		5 & 0.00    & 0.00     &   0.00  \\ \hline
	\end{tabular}
\end{table}

\begin{table}[htbp]
	\centering
	\caption{Illustrative Example. Estimated posterior mean (and standard deviation) of frequencies and square root of the power of the corresponding frequencies.   }
	\label{table:estimate_parameter_illustrative}
	\begin{tabular}{lcccc}
		\hline \\[-0.9em]
		\hline
		& $\omega_{\, 11}$                                          & $\omega_{\, 21}$                                          & $\omega_{\, 31}$                                          & $\omega_{\, 32}$                                                         \\ \cmidrule{2-5}
		True      & 0.0400                                                    & 0.0526                                                    & 0.0833                                                    & 0.1250                                                                                                 \\ [.2em]
		Estimated & \begin{tabular}[c]{@{}c@{}}0.0399\\ \footnotesize (8.8 $\cdot 10^{-6}$)\end{tabular} & \begin{tabular}[c]{@{}c@{}}0.0526\\ \footnotesize (6.3 $\cdot 10^{-6}$)\end{tabular} & \begin{tabular}[c]{@{}c@{}}0.0833\\ \footnotesize (9.6 $\cdot 10^{-6}$)\end{tabular} & \begin{tabular}[c]{@{}c@{}}0.1249\\ \footnotesize ($9.4 \cdot 10^{-6}$)\end{tabular}  \\ 
		&                                                           &                                                           &                                                           &                                                                                           \\
		& $A_{\, 11}$                                               & $A_{\, 21}$                                               & $A_{\, 31}$                                               & $A_{\, 32}$                                          \\ \cmidrule{2-5}
		True      & 1.131                                                    & 0.283                                                    & 1.414                                                     & 1.414                                                     \\ [.2em]
		Estimated & \begin{tabular}[c]{@{}c@{}}1.069 \\  \footnotesize (0.029)\end{tabular}  & \begin{tabular}[c]{@{}c@{}}0.281\\ \footnotesize (0.004)\end{tabular}  & \begin{tabular}[c]{@{}c@{}}1.380\\ \footnotesize (0.022)\end{tabular}  & \begin{tabular}[c]{@{}c@{}}1.367\\ \footnotesize (0.022)\end{tabular}  \\[.8em]
		\hline
	\end{tabular}
\end{table}

Table \ref{table:posterior_k_m_illustrative_1} (left panel) shows  that our estimation algorithm successfully detects the correct number of states in the sense that  a  model with $k = 3$  regimes has the highest posterior probability. In addition, our approach correctly identifies the right number of frequencies in each regime, as shown in Table \ref{table:posterior_k_m_illustrative_1} (right panel). Table \ref{table:estimate_parameter_illustrative}  displays the estimated posterior mean and standard deviation of the frequencies along with the square root of the power of the corresponding frequencies, where the results  are conditional on three estimated states and  the modal number of frequencies within each state.  Here, the power of each frequency $\omega_{\, jl}$ is summarized by the amplitude $A_{\, jl} =  \sqrt{\beta_{j l}^{\, (1)^{\,2}} + \beta_{j l}^{\, (2)^{\,2}}}$, namely the square root of the sum of squares of the corresponding linear coefficients (see, e.g., \citet{shumway2017time}). Our proposed method seems to provide a good match between true and estimated values for both frequencies and their power, for this example. We also show in Figure \ref{fig:illustrative_example} the state-specific estimated signal (Equation \eqref{eq:oscillatory}), and the estimated state sequence using the method of \citet{stephens2000dealing} (as a piecewise horizontal line). The rows of the estimated transition probability matrix were $\bm{\hat{\pi}}_1 = (0.9921, \, 0.0073, \, 0.0006), \, \bm{\hat{\pi}}_2 = (0.0005, \, 0.9956, \,0.0040)$ and $\bm{\hat{\pi}}_3 = (0.0051, \,0.0006, \, 0.9942)$.  The high probabilities along the diagonal reflect the estimated posterior mean of the self transition parameter $\hat{\rho} = 0.9860$, which is indeed centered around the true probability of self-transition.


\begin{figure}[htbp]
	\centering
	\centerline{\includegraphics[scale = 0.38]{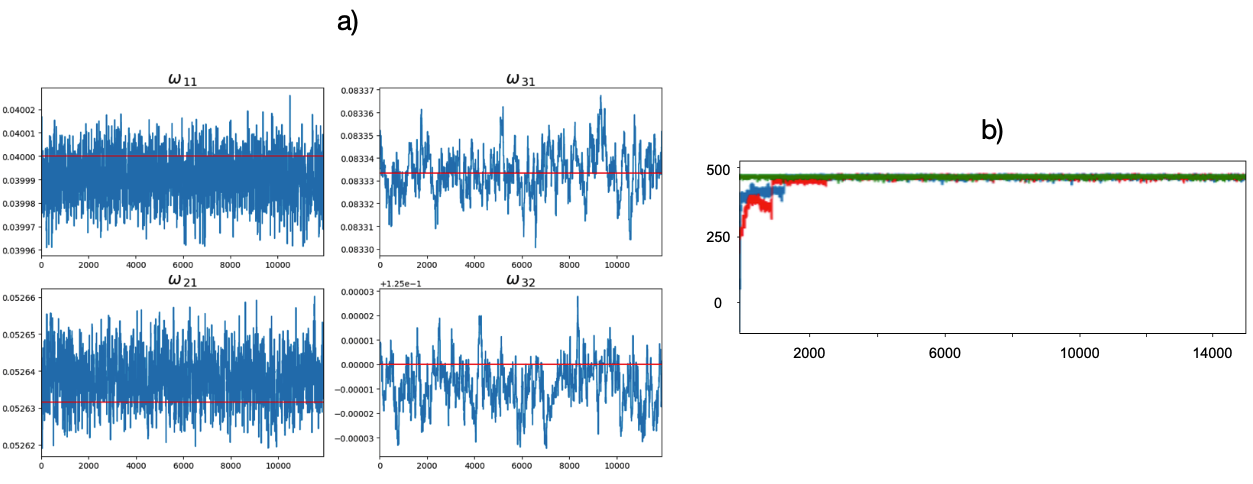}}
	\vspace{3mm}
	\caption{Illustrative Example. (a) Trace plots (after burn-in of 3000 updates) for posterior sample of frequencies, conditional on modal number of states and number of frequencies in each state; red lines correspond to true values of the frequencies. (b) Trace plots (including burn-in) of the likelihood for three Markov chains initialized at  different starting values (where the initial 100 updates are omitted from the graph) }
	\label{fig:traces_illustrativce}
\end{figure}

Diagnostics for verifying convergence were performed in several ways. For example, we observed that  the MCMC samples of the likelihood of the HMM reached a stable regime, while initializing the Markov chains from overdispersed starting values (see Figure \ref{fig:traces_illustrativce} (b)). This diagnostic might be very useful, for example, in determining the burn-in period. However, we note that it does not guarantee convergence since steady values of the log likelihood might be the result of a Markov chain being  stuck in some local mode of the target posterior distribution. The likelihood of an HMM with Gaussian emissions can be expressed as  
\begin{equation}
\mathcal{L} \, (\bm{z}, \bm{\pi}, \,  \bm{\theta} \, | \,  \bm{y} ) = p (z_1 \, | \, \bm{y}, \bm{\pi}, \bm{\theta}) \mathcal{N} ( y_1 \,  ; f_{\, 1\, z_{\,1}}, \, \sigma^{\, 2}_{z_{\,1}}) \prod_{t=2}^{T} p \, (z_{\, t} \, | \, z_{\, t-1}, \,  \bm{y}, \bm{\pi}, \bm{\theta}  \,) \mathcal{N} ( y_t \,  ; f_{\,t \,z_{\,t}}, \, \sigma^{\, 2}_{z_{\, t}}),
\end{equation}
where  $\mathcal{N} \, ( \,y_t; \, f_{j \, t}, \, \sigma^2_j )$ denotes the density of a Gaussian distribution with mean  $ f_{jt} = \bm{x}_t \, \big( \bm{\omega}_{ j} \big)^{\, '} \,  \bm{\beta}_{\, j} $ (as in Equation \ref{eq:oscillatory}) and variance $\sigma^2_j$, evaluated at $y_t$. Conditioned on the modal number of states, we also validated  convergence for the state-specific emission parameters by analyzing  trace plots and running averages of the corresponding MCMC samples, with acceptable results as each trace reached a stable regime. As an example, we show in Figure \ref{fig:traces_illustrativce} (a) trace plots (after burn-in) for the posterior values of the frequencies. 

Finally, while we notice that we  have not set  $K_\text{max}$ to a very large value, this choice has been made a \textit{posteriori} after we found that the estimation algorithm  assigned negligible probabilities to a large number of components. For example, in this simulation study, we initially set $K_{\text{max}} = 20$ and  ran the full estimation algorithm; after observing that the posterior probabilities for the number of distinct states were equal to zero for all the models with more than four hidden states, we re-ran the estimation algorithm with a smaller value for the HDP truncation, i.e. $K_{\text{max}}$ = 7,  obtaining the same correct results. This yielded some benefits from a computational perspective, in particular in terms of facilitating storage and memory access of the posterior sample, and also speeding up the relabelling algorithm developed by \cite{stephens2000dealing}. Furthermore, we notice that even in the case  where the maximum complexity of the model is set to $K_{\text{max}} = 7$, we are still dealing with a framework that assumes a  relatively high number of regimes (within a fairly complicated setting of oscillations in each state). Conventional models might not even be able to achieve satisfactory estimation performances when specifying such a large number of Markov modes.

\vspace{0.1cm}

\textit{Signal Extraction with Non-Gaussian Innovations:}    In many scientific experiments it may be of interest to extract the underlying  signal that generates the observed time series and HMMs can be used to this end. Here, we study the performance of our proposed approach in estimating the time-varying oscillatory signal $f_{\, j t \, }$ (Equation \ref{eq:oscillatory})  when the Gaussian assumption of $\varepsilon_t$ in Equation \eqref{eq:emission_distribution} is violated. In particular, we generated 20 time series, each consisting of 1024 observations from the same simulation setting introduced above, where the innovations were simulated from heavy-tailed $t-$distributions with 2, 3, 2 degrees of freedom for state 1,2,3, respectively. The  linear basis coefficients were chosen to be $\bm{\beta}_{\, 1 1} =  (3, 2)^{'}, \, \bm{\beta}_{\, 2 1} =  (1.2, 4.0)^{'}, \, \bm{\beta}_{\, 3 1} =  (1.0, 5.0)^{'}, \, \bm{\beta}_{\, 3 2} =  (4.0, 3.0)^{'}$.  As  a measure of performance we computed the (in-sample) mean squared error MSE = $ \frac{1}{1024}\sum_{t = 1}^{1024} ( f_{t \, z_{\,t}} - \hat{f}_{t \, z_{\,t}})^2$ between true and estimated signal and compared the proposed approach  with the method of \citet{hadj2018bayesian} referred to as AutoNOM (Automatic Nonstationary Oscillatory Modelling), which we believe is the state-of-the-art in extracting the signal of nonstationary periodic processes. Our proposed estimation algorithm was run with the same parameterization as above while AutoNOM was performed for 15,000 updates, 3,000 of which were discarded as burn-in, where we fixed 15 maximum number of change points and 5 maximum number of frequencies per segment. The prior means for the number of change-points and frequencies per segment are fixed at 2 and 1, respectively, and the minimum distance between change-points is set at 10. For both methodologies, the estimated signal was obtained by averaging across MCMC iterations. AutoNOM was run using the Julia software provided by the authors at \href{https://github.com/Beniamino92/AutoNOM}{\texttt{https://github.com/Beniamino92/AutoNOM}}.

\begin{figure}[htbp]
	\centering
	\includegraphics[height =8.0cm, width = 14.5cm]{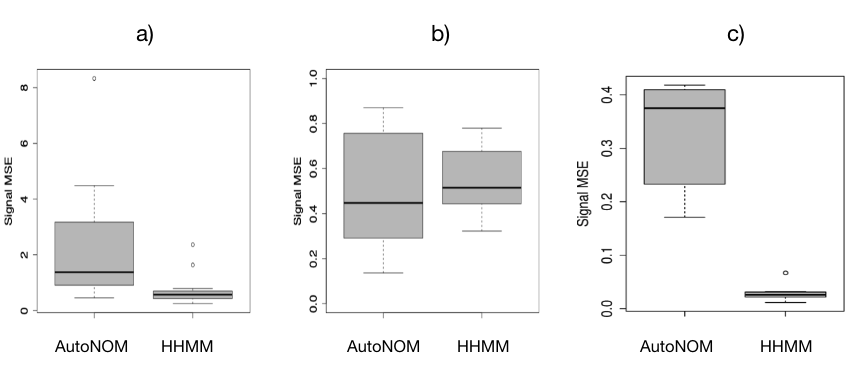}
	\caption{ Signal extraction with non-Gaussian innovations.  Boxplots of the MSE values for AutoNOM and our proposed HHMM when (a)  the data exhibit recurrent patterns; (b) the data do not exhibit recurrent patterns; (c) the innovations are skewed.} 
	\label{fig:t_dist_simul}
\end{figure}

Figure \ref{fig:t_dist_simul} (a) presents boxplots of the MSE values for AutoNOM and our proposed approach which will be referred to as HHMM (Harmonic Hidden Markov Model). It becomes clear that the estimates of the signal obtained using HHMM are superior to those obtained using AutoNOM. However, this result is not surprising as the two approaches make different assumptions. In particular, AutoNOM does not assume recurrence of a periodic behavior and hence needs to estimate the regime-specific modeling parameters each time it detects a new segment, while our HHMM has the advantage of using the same set of parameters whenever a particular periodic pattern recurs in the time series. Hence, we also compared the performance of the two approaches in extracting the signal (under non-Gaussian innovations) in a scenario where the time series do not exhibit recurrence.  Specifically, we generated 15 time series manifesting two change-points (where the oscillatory behavior corresponding to the three different partitions are parameterized as above) and computed the MSE between true and  estimated signal as we did in the previous scenario. The corresponding boxplots displayed in  Figure \ref{fig:t_dist_simul} (b) show that the two approaches seem to perform in  similar way, with AutoNOM being slightly more accurate than our HHMM. Moreover, we have  further examined a scenario where the noise is generated  from exponentially distributed random variables in order to introduce a large skew.  Here we simulated 15 time series from a three-state HMM as above where now the innovations corresponding to the three different states are generated from exponentially distributed random variables with rates 0.5, 1, and 0.2, respectively.  The draws from the exponential distribution are centered in such a way that they have mean zero, to avoid noise that takes on strictly positive values.  Figure \ref{fig:t_dist_simul} (c) presents boxplots of the MSE values for AutoNOM and our proposed HHMM, showing that our methodology seems to be superior to AutoNOM, in terms of extracting the signal when the innovations are skewed. We conclude that both approaches have their own  strengths. Our proposed procedure is superior to AutoNOM  in the sense that the additional HMM provides a framework for modeling and explicitly quantifying  the  switching dynamics and connectivity between different states. On the other hand,  AutoNOM is better  suited to scenarios where there are nonstationarities arising from singular change-points and the observed oscillatory processes evolve without recurrent patterns.

\subsection{Markov Switching Autoregressive Process}
We now investigate the performance of our approach in detecting time-changing periodicities in a scenario where the data generating process shows large departures from our modeling assumptions. The HMM hypothesis which assumes  conditionally independent observations given the hidden state sequence, such as the one formulated in Equation \eqref{eq:emission_distribution}, may sometimes be  inadequate in expressing the temporal dependencies occurring in some phenomena. A different class of HMMs  that relaxes this assumption is given by the \textit{Markov switching autoregressive process}, also referred to as the AR-HMM \citep{juang1985mixture, albert1993bayes, fruhwirth2006finite}, where an AR process is associated with each state. This model is used to design autoregressive dynamics for the emission distributions  while allowing the state transition mechanisms  to follow a discrete state Markov chain.  

\begin{figure}[htbp]
	\centering
	\includegraphics[height =8.8cm, width = 7.8cm]{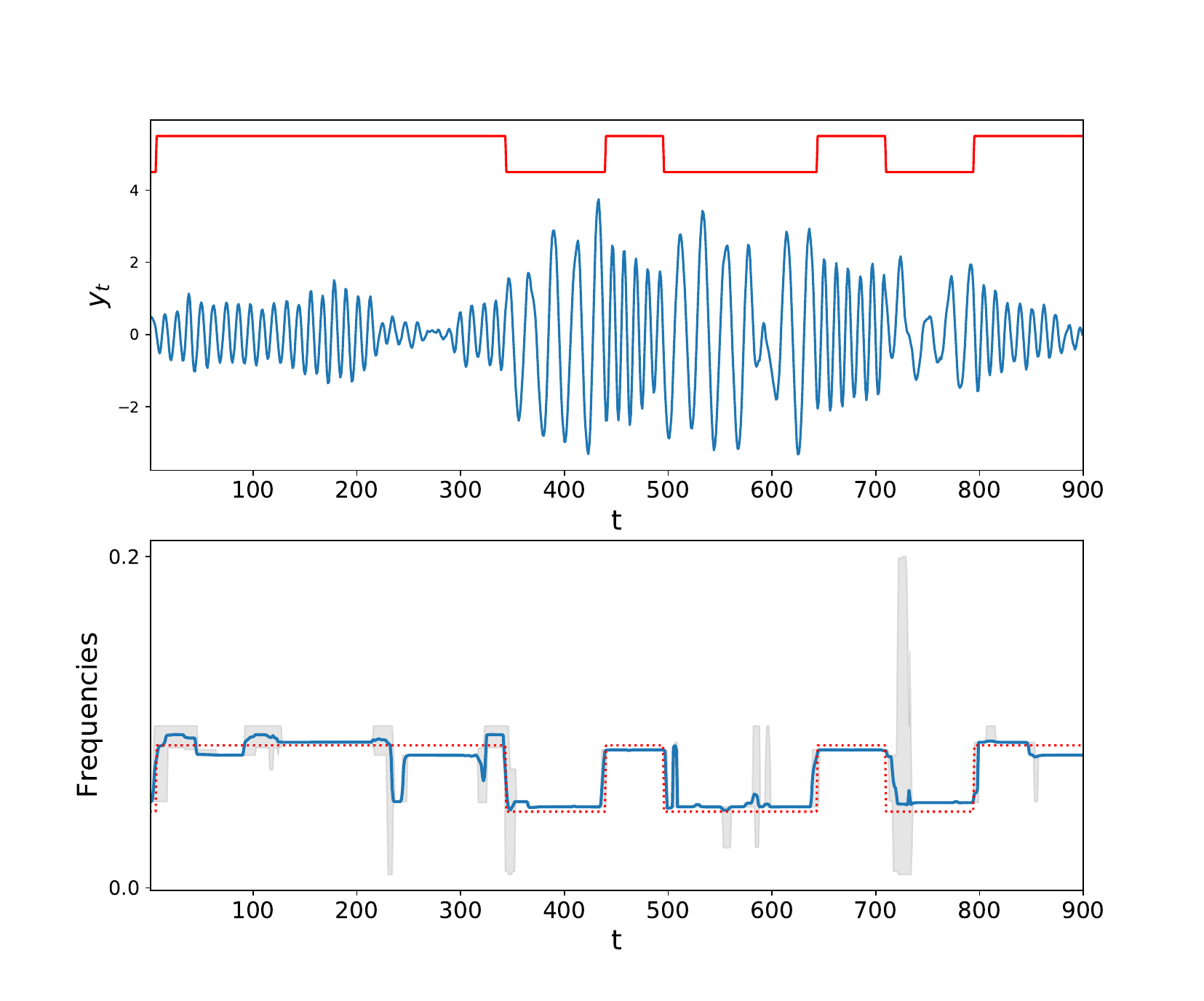}
	\includegraphics[height =8.4cm, width = 5.6cm]{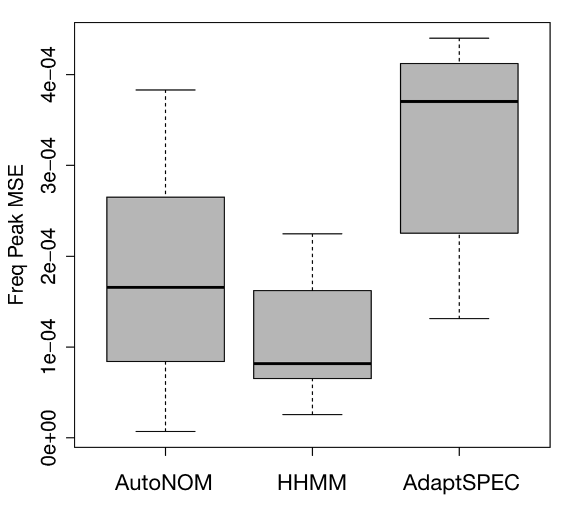}
	\caption{(Top) A realization from model \eqref{eq:AR_HMM}, where the piecewise horizontal line represents the true state sequence. (Bottom) True time varying frequency peak  (dotted red line) and the estimate provided by our proposed approach (solid blue line) where we highlight a 95\% credible interval obtained from the posterior sample. (Right) Boxplots of the MSE values for AutoNOM, our proposed HHMM and AdaptSPEC.}
	\label{fig:AR_HMM_study}
\end{figure}

We generated $T = 900$ observations from an AR-HMM with two hidden states and  autoregressive order fixed at $p = 2$, that is 
\begin{equation} \label{eq:AR_HMM}
\begin{split}
z_{\,t} \, \sim& \, \, \bm{\pi}_{z_{\,t-1}}, \\
y_{\,t} =& \sum_{l=1}^{p} \psi^{\, (z_t)}_{\,l} y_{\, t-l} + \varepsilon^{\, (z_t)},
\end{split}
\end{equation}
where  $\bm{\pi}_1 = (0.99, 0.01)$ and $\bm{\pi}_2 = (0.01, 0.99) $. The AR parameterization  $\bm{\psi}^{\, (1)} = (1.91, \, -0.991)$ and $\bm{\psi}^{\, (2)} = (1.71, \, -0.995 )$  is  chosen in such a way that the state-specific spectral density functions display a pronounced peakedness. Furthermore,  $\varepsilon_t^{(1)} \stackrel{iid}{\sim} \mathcal{N}(0, 0.1^{\,2})$ and $\varepsilon_t^{(2)} \stackrel{iid}{\sim} \mathcal{N}(0, 0.05^{\,2})$. A realization from this model is shown in Figure \ref{fig:AR_HMM_study} (top) as a blue solid line. Our proposed estimation algorithm was run for 15,000 iterations 5,000 of which are used as burn-in. At each iteration, we performed 2 reversible-jump MCMC updates for each instantiated set of emission parameters. The rate of the Poisson prior for the number of periodicities is fixed at $10^{-1}$ and the corresponding truncation level $d_{\text{max}}$ was fixed to 3. The maximum number of states  $K_{\text{max}}$ was set equal to 10 whereas the rest of the hyperparameters are specified as in Section \ref{sec:illustrative_example_simul}.  Our procedure seems to overestimate the number of states, as a model with 8 regimes had the highest posterior probability $\hat{\pi} \, (\, k = 8 \, | \, \bm{y}) = 97\%$. However, this is not entirely unexpected as a visual inspection of the realization  displayed in Figure \ref{fig:AR_HMM_study} (top)  suggest more than two distinct spectral patterns in the sense that the phases,  amplitudes  and  frequencies appear to  vary  stochastically  within  a  regime.  
Figure \ref{fig:AR_HMM_study} (bottom) shows the estimated  time-varying frequency peak  along with a 95\% credible interval obtained from the posterior sample. The  estimate was  determined by first selecting the dominant frequency (i.e. the frequency with the highest posterior power) corresponding to  each observation and then averaging the frequency estimates over MCMC iterations. While our approach identifies a larger number of states  when the data were generated  from an AR-HMM we note that the data generating process are very different from the assumptions of our model and  the proposed procedure still provides  a reasonable  summary of the underlying time changing spectral properties observed in the data. Furthermore, by setting the truncation level $K_{max}$ equal to 2, we retrieve the true transition probability matrix that generates the switching dynamics between the two different autoregressive patterns, as the vectors of transition probabilities obtained using our estimation algorithm are $\hat{\bm{\pi}}_1 = (0.99, 0.01)$ and $\hat{\bm{\pi}}_2 = (0.98, 0.02).$

In addition, we simulated 10 time series from  model \eqref{eq:AR_HMM} and computed the mean squared error MSE = $\frac{1}{900} \sum_{t=1}^{900} ( \omega_t - \hat{\omega}_t)$ between the true time-varying frequency peak $\omega_t$ and its estimate $\hat{\omega}_t$  for the proposed approach, AutoNOM and the procedure of \citet{rosen2012adaptspec}, referred to as AdaptSPEC (Adaptive Spectral Estimation).  For both AutoNOM and AdaptSPEC, we ran the algorithm for 15,000 MCMC iterations (5,000 of which were used as burn-in), fixed the maximum number of change-points at 15 and set the minimum distance between change-points to 30. The number of  spline basis functions for AdaptSPEC is set to 10. AutoNOM is performed using a Poisson prior with rate $10^{-1}$ for both  number of frequencies and number of change-points. AdaptSPEC was performed using the R package \textit{BayesSpec} provided by the authors. Boxplots of the MSE values for the three different methodologies are displayed in Figure  \ref{fig:AR_HMM_study} (right), showing that our proposed HHMM seems to outperform the other two approaches in detecting the time-varying frequency peak, for this example. However, our procedure finds some very short sequences (such as in Figure \ref{fig:AR_HMM_study} (bottom) for  $t \approx 200, 500, 700$) demonstrating that the sticky parameter might not  always be adequate enough in capturing the correct temporal mode persistence of the latent state sequence. AutoNOM and AdaptSPEC are less prone to this problem as both methodologies are able to specify a minimum time distance between change-points; though, we  acknowledge  that this constraint might not be optimal when the observed data exhibit relatively rapid changes. We also notice that, not surprisingly, the estimates of the time-varying frequency peak obtained using  AutoNOM and our HHMM, which are based on a line-spectrum model, are both superior than the ones obtained via the smoothing spline basis of AdaptSPEC, which is built upon a continuous-spectrum setting; this is consistent with the findings in \citet{hadj2018bayesian}. However, it is important to keep in mind that,  while AutoNOM and AdaptSPEC allow to retrospectively analyse the spectral changing properties of a process from an exploratory angle, unlike our proposed HHMM, they do not quantify a probabilistic mechanism for the recurrence of periodic dynamic patterns.

\section{Analysis of the Airflow Trace Data} \label{sec:case_study}
The airflow trace shown in Figure \ref{fig:human_breathing_trace_intro}  was  collected from a human over a time span of 5.5 minutes of continuous 
breathing and measured via a facemask attached to a pressure transducer. Airflow pressure signals were amplified using the NeuroLog system connected to a 1401 interface and acquired on a computer using \textit{Spike2} software (Cambrdige Electronic Design). The data are  sampled at rate of 4 Hertz, i.e., 4 observations per second,
for a total of 1314 data points.   The air flow data was captured and amplified only, and no pre-processing was carried out afterwards. We notice that the signal is clean since the pressure transducer is sensitive enough to pick up breathing even from a mouse in a plethysmograph, and it was attached to a medical face-mask directly in front of the mouth. Therefore, the signal to noise ratio is extraordinarily high.

We fitted our HHMM to the time series displayed in Figure \ref{fig:human_breathing_trace_intro} for 200,000 iterations, 125,000 of which were discarded as burn-in, where at each iteration, we carried out 10 reversible-jump MCMC updates for each instantiated set of emission parameters.  The truncation level $K_{\text{max}}$ was set to 10, whereas the maximum number of frequencies  per state $d_{\text{max}}$ was  fixed to 3. With respect to the  harmonic regression part of the model, we specified the prior for the innovations $\sigma^2_j$ as Inverse-Gamma $(3.11, 3.17)$, so that it is centered at the empirical variance $1.5$ and has a standard deviation of $\sqrt{2.5}$. The Poisson prior for the number of frequencies $d_j$ is chosen equal to $10^{-2}$, to favour models with a small number of components, and the prior on the frequencies $\omega_{j,l}$ is set informative as Uniform(0, $\phi_{\omega})$, where $\phi_{\omega} = 0.3$ was selected by looking at the raw periodogram of the data and noting that the power of the frequencies was approximately zero for any values larger than 0.3. Finally, we specified weakly informative prior on the linear basis coefficients $\bm{\beta}_j$  as  $\mathcal{\bm{N}}_{2d_j} (\, \bm{0}, \, \sigma^2_{\beta} \, \bm{I}\, )$, where the prior variance $\sigma^2_\beta = 4$ is chosen in now such a way that the mass of the prior is concentrated in reasonable regions based on the data. Regarding the part of the model relative to the HDP, we specified for both  concentration parameters $\eta + \kappa$ and $\gamma$  a weakly informative hyperprior  Gamma $(1, 0.01)$, so that the corresponding priors for the base measures favour a DP model with a small number of components (see stick breaking construction, Equations \eqref{eq:stick_breaking_1} and \eqref{eq:stick_breaking_2}). The prior on the self transition proportions $\rho$ is specified informative as Beta $(10^{\,3}, 1)$, so that we  force high probability of self transitions.

\begin{figure}[htbp]
	\centering
	\centerline{\includegraphics[scale = 0.47]{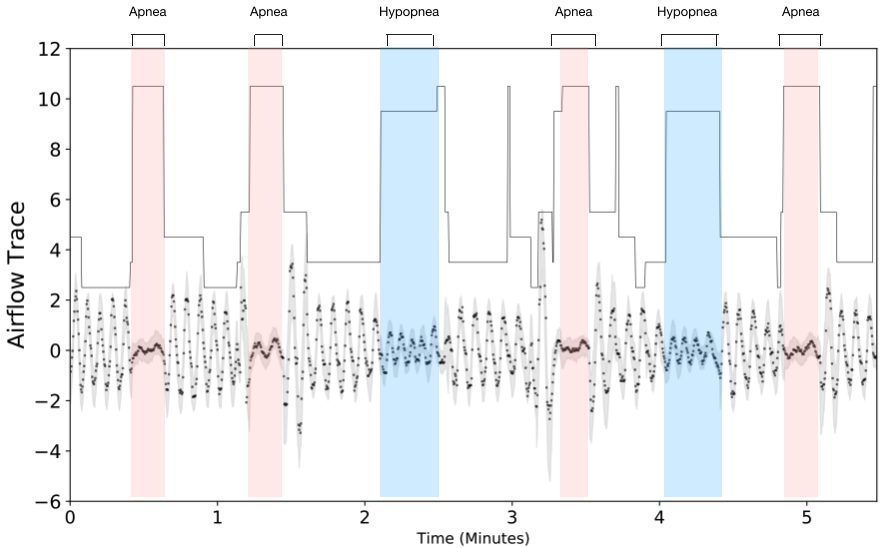}}
	\caption{Case Study. Dots represent the airflow trace collected over a period of five and half minutes of continuous breathing. The grey lines represent draws from the estimated posterior predictive. The piecewise horizontal line corresponds to the estimated state sequence where we highlight the states corresponding to estimated apnea (red) and hypopnea (blue), while  reporting the ground truth at the top of the plot. }
	\label{fig:human_breathing_trace}
\end{figure}

The posterior distribution over the number of states  had a mode at 6, with posterior probabilities $\hat{\pi} \, (\, k = 6 \, | \, \bm{y}) = 94\%$,  $\hat{\pi} \, (\, k = 7 \, | \, \bm{y}) = 5\%$ and  $\hat{\pi} \, (\, k = 8 \, | \, \bm{y}) = 1\%$. Indeed, it is conceivable that the state corresponding to \textit{normal} breathing (i.e. neither apnea or hypopnea) may exhibit more than one distinct periodic pattern, which further justifies the need to use a nonparametric HMM.  \cite{paz2013acute} reported at least 13 forms of breathing patterns including forms of apnea. Figure \ref{fig:human_breathing_trace} shows the  estimated hidden state sequence (piecewise horizontal line), where we highlight our model estimate for apnea state (red) and hypopnea state (blue) while  reporting the ground truth at the top of the plot. We have also included  a  posterior predictive graphical check consisting of the airflow trace alongside 20 draws from the estimated posterior predictive \citep{gelman2014bayesian}, where the latter is obtained by first drawing a sample path, and then, conditioned on the hidden state sequence, the predicted values are simulated from the appropriate emission distributions. Our model seems to be able to capture the underlying signal that characterizes  this time series. Conditional on the modal number of regimes, the number of periodicities belonging to apnea and hypopnea had a posterior mode at 3 and 2, respectively. Conditional on the modal number of frequencies, Table \ref{table:estimate_parameter_apnea_hypopnea} displays the posterior mean and standard deviation of periodicities (in seconds) and powers that characterize the two states classified as apnea and hypopnea, showing that apnea instances seem to be characterized by larger periods and lower amplitude than hypopnea.

\vspace{0.7cm}

\begin{table}[htbp]
	\centering
	\caption{Case study. Posterior mean and standard deviation of frequencies and corresponding powers that characterize the two states classified as apnea and hypopnea. }
	\label{table:estimate_parameter_apnea_hypopnea}
	\begin{tabular}{cclcc}
		\hline \\[-0.9em]
		\hline
		
		\multicolumn{2}{c}{\textbf{Apnea}}                                                                                                &  & \multicolumn{2}{c}{\textbf{Hypopnea}}                                                                                            \\
		Freq                                                     & Power                                                       &  & Freq                                                      & Power                                                     \\ [.1em] \cmidrule{1-2}  \cmidrule{4-5}
		\begin{tabular}[c]{@{}c@{}}0.0159\\ \footnotesize (6.66$\cdot 10^{-5}$)\end{tabular} & \begin{tabular}[c]{@{}c@{}}0.1376\\ \footnotesize (1.81$\cdot 10^{-2}$)\end{tabular} &  & \begin{tabular}[c]{@{}c@{}}0.0455\\ \footnotesize (1.29$\cdot 10^{-4}$)\end{tabular}  & \begin{tabular}[c]{@{}c@{}}0.261\\ \footnotesize (3.37$\cdot 10^{-2}$)\end{tabular} \\[1.2em]
		\begin{tabular}[c]{@{}c@{}}0.0353\\ \footnotesize (3.73$\cdot 10^{-5}$)\end{tabular} & \begin{tabular}[c]{@{}c@{}}0.2153\\ \footnotesize (1.87$\cdot 10^{-2}$)\end{tabular} &  & \begin{tabular}[c]{@{}c@{}}0.0542\\ \footnotesize (4.72$\cdot 10^{-5}$)\end{tabular}  & \begin{tabular}[c]{@{}c@{}}0.620\\ \footnotesize (3.4$\cdot 10^{-2}$)\end{tabular} \\ [1.2em]
		\begin{tabular}[c]{@{}c@{}}0.0379\\ \footnotesize (3.89$\cdot 10^{-5}$)\end{tabular}                                                          & \begin{tabular}[c]{@{}c@{}}0.2065\\ \footnotesize (1.97$\cdot 10^{-2}$)\end{tabular}                                                           &  & - & -
		\\ [.2em] \hline
	\end{tabular}
	
\end{table}

\newpage

Our estimation algorithm detected all known apnea and hypopnea instances.  In order for them to qualify as a clinically relevant obstructive event they must have a minimum length of 10 seconds \citep{berry2017aasm}. Thus, we only highlight the clinically relevant instances in Figure \ref{fig:human_breathing_trace}, discarding sequences of duration less than 10 seconds. We also detected a \textit{post sigh} apnea (after the third minute) which is a normal phenomenon to observe in a breathing trace and hence should not count as  a disordered breathing event. Again, such an event after a sigh can be identified as a sigh is characterized by an amplitude  which is always higher than any other respiratory event and hence can  be easily detected. Subtracting the number of sighs from the total number of apneas/hypopneas results in a measure of all apneas of interest without the confounding data from post sigh apneas. A common score to indicate the severity of sleep apnea is given by the Apnea-Hypopnea Index (AHI) which consists of the number of apneas and hypopneas per hour of sleep \citep{ruehland2009new}. Our proposed approach provides a realistic estimate of the total number of apnea and hypopnea instances recurring in this case study.  While an essential aim of this paper is to detect apnea instances retrospectively, which is currently a time consuming and demanding task as it is performed by eye, we have also  investigated the out-of-sample predictive performance of our proposed approach in Supplementary Material \textcolor{blue}{E}.


\section{Summary and Discussion} \label{sec:conclusions}
In this paper we developed a novel HMM approach that can address the challenges of modeling periodic phenomena whose behavior switches and recurs dynamically over time. The number of states is assumed unknown as well as their relevant periodicities which may
differ over the different regimes since each regime is represented by a different periodic pattern. To address flexibility in the number of states, we assumed a sticky HDP-HMM that penalises rapid changing dynamics of the process and provides effective control over the switching rate. The variable dimensionality with respect to the number of frequencies that specifies the different states is tackled by a reversible-jump MCMC algorithm.

While being 
noticeably flexible, the model proposed in this article is developed in a framework that is still computationally accessible. Naturally, an alternative strategy would involve fitting several finite-state HMMs and then performing in a second stage model selection  by means of the marginal likelihood \citep{kass1995bayes}.  Nevertheless, reliably approximating this quantity from the posterior sample is not straightforward, though several techniques have been proposed in the literature to overcome this burden, such as sequential Monte Carlo  (SMC, \citealt{jasra2008interacting}), population MCMC (PMCMC, \citealt{liang2001real, jasra2007population}) or bridge sampling \citep{meng1996simulating, meng2002warp}, where we refer the reader to \citet{zhou2016toward} for an excellent summary of these developments. However, these methods involve algorithms that are often computationally challenging and difficult to implement efficiently, especially within our modeling framework.


We illustrated the use of our approach in a case study relevant to respiratory research, where our methodology was able to  identify recurring  instances of sleep apnea in human
breathing traces. Despite the fact that here we have focused on the detection of apnea instances, our proposed methodology provides a very flexible and general framework to analyze different breathing patterns.  A question of interest is whether similar dynamical patterns can be identified across a heterogeneous patient cohort, and be used for the prognosis of patients' health and progress. The growth of information and communication technologies permits new advancements in the health care system to facilitate support in the homes of patients in order to proactively enhance their health and well-being. We believe that our proposed HMM approach has the potential to aid the iterative feedback between clinical investigations in sleep apnea research and practice with computational, statistical and mathematical analysis.

As pointed out by a referee, apnea states have certain features, such as low amplitude and low frequency behaviour, that may suggest that assuming symmetry among the parameters in their prior distribution might not be an ideal modeling approach.  However,  as we discussed in this article, it is plausible that normal breathing is exhibiting more than one distinct periodic pattern. While it would be of interest to integrate  prior knowledge into the model to fully remove permutations of labelling of the HDP-HMM mixture components, we believe it would not be trivial to fully characterize in an identifiable way  the different states corresponding to normal breathing. These are early days for such data to be analyzed in this way and we are at the beginning of being able to construct a catalogue of more informative priors that might help this type of analysis in future.

Although both parametric and nonparametric HMMs have been shown to be good models in addressing learning challenges in time series data, they have the drawback of limiting the \textit{state duration distribution}, i.e., the distribution for the number of consecutive time points that the Markov chain spends in a given state, to a geometric form \citep{ephraim2002hidden}. In addition, the self-transition bias of the sticky HDP-HMM  used to increase temporal state persistence is shared among all states and thus  does not allow for inferring state-specific duration features. In our application, learning the duration structure of a specific state  may be  of interest to health care providers, for example,  in assessing the severity of sleep apnea. Future work will address extending our approach to a hidden semi-Markov model (HSMM) setting \citep{guedon2003estimating,yu2010hidden,johnson2013bayesian}, where the generative process of an HMM is augmented by introducing a random state duration time which is drawn from some state-specific distribution when the state is entered. However, this increased flexibility in  modeling the state duration  has the cost of  increasing substantially the computational effort to compute the likelihood: the message-passing procedure for HSMMs requires $\mathcal{O} (T^2 K + T K^2)$ basic computations for a time series of length $T$ and number of states $K$, whereas the corresponding forward-backward algorithm for HMMs requires only $\mathcal{O}(T K^2)$.



\section*{Acknowledgements} 

We wish to thank Maxwell Renna, Paul Jenkins and Jim Griffin for their insightful and valuable comments. The work presented in this article was developed as part of the first author's Ph.D. thesis at the University of Warwick and he is currently affiliated with the  Department of Statistics at Rice University. B. Hadj-Amar was supported by the Oxford-Warwick Statistics Programme (OxWaSP) and the Engineering and Physical Sciences Research Council (EPSRC), Grant Number EP/L016710/1.  R. Huckstepp was supported by the Medical Research Council (MRC), Grant Number MC/PC/15070.

\vspace{1.0cm}

\begin{center}
	SUPPLEMENTARY MATERIAL
\end{center}

We provide supplemental material to the manuscript. Section \textcolor{blue}{A} contains additional details about the sampling scheme for updating the emission parameters via  reversible-jump MCMC steps, and in Section  \textcolor{blue}{B} we present the sampling scheme for drawing the HMM parameters within the Chinese restaurant franchise framework. Section \textcolor{blue}{C} gives the parameterization of the simulation setting presented in Section \ref{sec:illustrative_example_simul}. Section \textcolor{blue}{D} provides further diagnostics about our MCMC sampler and in Section \textcolor{blue}{E} we investigate the out-of-sample predictive performance of the proposed approach in the air flow case study. Julia code that implements our  proposed approach is also available at \href{https://github.com/Beniamino92/HHMM}{\texttt{https://github.com/Beniamino92/HHMM}}.

\bibliography{Biblio}       

\end{document}